\documentclass[twocolumn]{autart}    

\usepackage{graphicx}          

\usepackage[sort,numbers]{natbib}        

 \usepackage{amssymb, amsmath, color}

\usepackage{enumerate}

\newcommand{\QED}{\hfill~$\blacksquare$  }

\newcommand{\mbb}[1]{\mathbb #1}

\newcommand{\mcl}[1]{\mathcal #1}

\newcommand{\OCP}[1]{\operatorname{\mathcal{OCP}}_T(#1)}
\newcommand{\interior}{{\operatorname{int}}}
\newcommand{\inte}[1]{\interior{ (\mcl{#1})}}

\newcommand{\Rnx}{\mbb{R}^{n_x}}

\newcommand{\Rnu}{\mbb{R}^{n_u}}

\newcommand{\R}{\mbb{R}} 

\newtheorem{ass}{Assumption}
\newtheorem{defi}{Definition}
\newtheorem{coro}{Corollary}

\newtheorem{lemm}{Lemma}
\newtheorem{stat}{Statement}
\newtheorem{rema}{Remark}
\begin{document}

\begin{frontmatter}

\title{On Turnpike and Dissipativity Properties of Continuous-Time Optimal Control Problems\thanksref{footnoteinfo}} 

\thanks[footnoteinfo]{A preliminary version of this paper was presented at  CDC 2014 \cite{epfl:faulwasser14e}.  
Corresponding author: Timm Faulwasser.
}

\author[KIT]{Timm Faulwasser}\ead{ timm.faulwasser@kit.edu},    
\author[UCSB]{Milan Korda}\ead{milan.korda@engineering.ucsb.edu},  
\author[EPFL]{Colin N. Jones}\ead{colin.jones@epfl.ch},  
\author[EPFL]{Dominique Bonvin}\ead{dominique.bonvin@epfl.ch}               

\address[KIT]{Institute for Applied Computer Science, Karlsruhe Institute of Technology, D-76131 Karlsruhe, Germany}
\address[UCSB]{Department of Mechanical Engineering, University of California, Santa Barbara}
\address[EPFL]{Laboratoire d'Automatique, \'Ecole Polytechnique F\'ed\'erale de Lausanne, CH-1015 Lausanne, Switzerland}  

\begin{keyword}                           
dissipativity, turnpike properties, converse turnpike results, optimal operation at steady state, optimal control, economic MPC             
\end{keyword}                             

\begin{abstract}                          
This paper investigates the relations between three different properties, which are of importance in optimal control problems: dissipativity of the underlying dynamics with respect to a specific supply rate, optimal operation at steady state, and the turnpike property. We show in a  continuous-time setting that if along optimal trajectories   a strict dissipation inequality is satisfied, then this implies optimal operation at this steady state and the existence of a turnpike at the same steady state. 
Finally, we establish novel converse turnpike results, i.e., we show that the existence of a turnpike at a steady state implies optimal operation at this steady state and dissipativity with respect to this steady state. 
We draw upon a numerical example to illustrate our findings.  
\end{abstract}
\end{frontmatter}

The notion of turnpike property of an optimal control problem (OCP)---introduced in \cite{Dorfman58} in the late 1950s---
is used to describe the phenomenon that in many finite-horizon OCPs the optimal solutions for different initial conditions approach a neighborhood of the best steady state, stay within this neighborhood for some time, and might leave this neighborhood towards the end of the optimization horizon. Turnpike phenomena have been observed in different types of OCPs: with/without terminal constraints  \cite{Carlson91, Clarke13a,  Trelat15a} and with/without discounted cost functionals \cite{Wuerth09, Zaslavski14a, Gurman04a}. 
Turnpikes have received widespread interest in the context of optimal control of economics \cite{McKenzie76, Carlson91}. The works by \cite{Anderson87a, Wilde72a, Rao99a, Sahlodin15} show how turnpike phenomena can be used to approximate solutions of OCPs with long horizons appearing in applications.\footnote{We remark that occasionally turnpike phenomena are denoted by varying names: \cite{Anderson87a, Wilde72a} refer to turnpikes as a \textit{dichotomy of optimal control problems}, while \cite{Rao99a} uses the phrase \textit{hypersensitive optimal control problems}.} Turnpikes also appear in OCPs arising in economic MPC formulations \cite{Wuerth09, epfl:faulwasser15g, epfl:faulwasser15a, kit:faulwasser17a, Gruene13a}.
%
%
%
      Recently, \cite{Stieler14a,Gruene16a} discussed different aspects of turnpike phenomena in a discrete-time setting with constraints and in a continuous-time setting without constraints \cite{Trelat15a}.
Taking into account the large number of publications on turnpike phenomena, it is quite surprising that only very few works state a precise definition of turnpike properties, see \cite{Zaslavski14a, Stieler14a}. 
Often, turnpike results for specific OCPs are proven without a rigorous definition of the turnpike property itself \cite{McKenzie76, Carlson91, Clarke13a, Gurman04a}. 
 While such an approach simplifies the construction of many turnpike results, it hinders establishing \textit{converse turnpike theorems}. 
    \vspace*{-3mm} 
 
  The main goal of this paper is to analyze the relation between three different properties that arise in the context of finite-horizon continuous-time OCPs: system dissipativity with respect to a specific supply rate (which depends on the cost function of the OCP), optimal operation at steady state, and the existence of a turnpike at that steady state. 
    Recently, a related  discrete-time analysis has been presented under the assumptions of local controllability of the turnpike and turnpike-like behavior of nearly optimal solutions \cite{Gruene16a}.
        The present paper takes a different route by avoiding such assumptions in the continuous-time case. Its contributions are as follows: 
While the preliminary version of this paper \cite{epfl:faulwasser14e} discussed \textit{state} turnpikes, we extend these results and provide a framework for the definition of different turnpike properties of OCPs, i.e., we suggest to distinguish \textit{state, input-state}, and \textit{extremal turnpikes} of OCPs. 
Our main contribution are novel converse turnpike results that  require neither local controllability of the turnpike nor turnpike-like behavior of nearly optimal solutions as in \cite{Gruene16a}. In particular, we show that the existence of a turnpike implies optimal operation at steady state; we prove that exactness of turnpikes implies dissipativity, whereby exactness of a turnpike means that the optimal solutions are at the turnpike steady state for some parts of the optimization horizon; and we show that under mild local assumptions on the cost function of the OCP, the existence of a turnpike implies satisfaction of a strict dissipation inequality along optimal solutions. 

 %
   The remainder of this paper is structured as follows: Section \ref{sec:prelim} introduces a formal definition of turnpike and dissipativity properties as well as the definition of optimal operation at steady state. Section \ref{sec:D} discusses implications of dissipativity. Section \ref{sec:SS} investigates the relation between optimal operation at steady state and dissipativity, while Section \ref{sec:T} presents converse turnpike results. 
   To demonstrate how some of our conditions can be verified, we draw upon the numerical example of a chemical reactor in Section \ref{sec:example}.

\section{Preliminaries and Problem Statement} \label{sec:prelim}
We briefly recall the notions of optimal operation at steady state, dissipativity with respect to a steady state, and turnpike properties of OCPs. 

\subsection{Optimal Steady-State Operation}
We consider the nonlinear system given by
\begin{align}
\dot x &= f(x,u), \quad x(0) = x_0,  \label{eq:sys_def} 
\end{align}
where the states $x \in \Rnx$ and the inputs $u \in \Rnu$ are constrained to lie in the compact sets 
$\mcl{X} \subset \Rnx$ and $\mcl{U} \subset \Rnu$.  We assume that the vector field $f:\Rnx\times\Rnu \to \Rnx$ is Lipschitz on~$\mathcal{X}\times \mathcal{U}$. A solution to  \eqref{eq:sys_def}, starting at $x_0$ at time $0$, driven by the input $u:[0,\infty) \to \mcl{U}$, is denoted as $x(\cdot, x_0, u(\cdot))$.

   \vspace*{-2mm} Consider the maximal control-invariant set $\mcl{X}_0 \subseteq \mcl{X}$ given by
\begin{multline}
\mcl{X}_0 = \left\{x_0 \in \mcl{X}\,|\,\exists\, u(\cdot)\in\mcl{L}([0,\infty), \mcl{U}):\right. \\
\left.\forall t \geq 0 ~x(t,x_0, u(\cdot)) \in \mcl{X}\right\},
\end{multline}
where $\mcl{L}([0,\infty), \mcl{U})$ denotes the class of measurable functions on $[0,\infty)$ taking values in the compact set $\mcl{U}\subset \Rnu$.
This set is the largest subset of $\mcl{X}$ that can be made positively invariant via a control $u(\cdot)$. Here, we assume that $\mathcal{X}_0 \not = \emptyset$. Furthermore, consider a finite-horizon OCP that aims at minimizing
the objective functional
\begin{equation} \label{eq:J}
J_T(x_0, u(\cdot)) = \frac{1}{T}\int_{0}^{T} F(x(t), u(t))\,\mathrm{d}t,
\end{equation}
where $F:\mcl{X} \times \mcl{U}\to \R$ is the cost function, and $T$ is the optimization horizon.  
We assume that $F$ is Lipschitz on $\mcl{X} \times \mcl{U}$.
The OCP reads 
\begin{subequations} \label{eq:OCP}
\begin{align}
\underset{u(\cdot) \in  \mcl{L}([0,T], \mcl{U})}{\inf} & \,J_T(x_0, u(\cdot))\\
\hspace*{-1cm}\textrm{ subject to} \nonumber \\ 
\dot x(t) & = f(x(t), u(t)), \quad x(0) = x_0 \\
\forall t \in  [0,T]: \quad x(t) &\in \mcl{X}, \quad u(t) \in \mcl{U}.
\end{align}
\end{subequations}
 The pair $\left(x(\cdot, x_0, u(\cdot)), u(\cdot)\right)$ is called admissible if  $u(\cdot) \in  \mcl{L}([0,T], \mcl{U})$ and if there exists a corresponding absolutely continuous solution $x(\cdot, x_0, u(\cdot))$, which  satisfies $x(t, x_0, u(\cdot)) \in \mcl{X}$  for all $t \in [0,T]$. 
 An optimal solution to \eqref{eq:OCP} is denoted by $u^\star(\cdot)$ and the corresponding state trajectory is written as $x^\star(\cdot, x_0, u^\star(\cdot))$.\footnote{Here, we assume for simplicity that the optimal solution exists and is attained. We refer to \cite{Vinter10, Lee67} for conditions ensuring the existence of optimal solutions to OCP~\eqref{eq:OCP}.}

\underline{Notational remarks:}  
We denote the dependence of optimal solutions to \eqref{eq:OCP} on the initial condition $x_0$ and the horizon length $T$  by writing 
$\OCP{x_0}$. Whenever it is convenient, input-state pairs are written as $z= (x,u)^T$ and the combined input-state constraints are written as $\mcl{Z} = \mcl{X}\times\mcl{U}$.
Throughout this paper, we use the superscript $\bar\cdot$ to denote a variable at steady state. Hence, we have $f(\bar z) = f(\bar x, \bar u) = 0$.
The set of admissible steady-state pairs is denoted as 
\begin{equation*}
\bar{\mcl{Z}}:= \left\{\bar z \in\mcl{Z}~|~0 = f(\bar z)\right\}.
\end{equation*}
Admissible trajectory pairs of $\OCP{x_0}$ are abbreviated by $z(\cdot, x_0) = (x(\cdot, x_0, u(\cdot)), u(\cdot))^T$.
For any function $\varphi$ with domain $\mbb{R}^{n_x+n_u}$ we write
\[
\varphi\left(z^\star(t, x_0)\right) := \varphi\left(x^\star(t, x_0, u^\star(\cdot, x_0)), \,u^\star(t, x_0)\right).
\]
While $\OCP{x_0}$ aims at optimizing the transient performance of system \eqref{eq:sys_def}, one can as well ask for the best stationary operating conditions. 
These conditions are given by the following steady-state problem: 
\begin{align} \label{eq:OPT_ss}
\underset{\bar z\,\in\,\mathbb{R}^{n_x+n_u}}{\inf} \, F(\bar z)\qquad \textrm{subject to } \bar z  \in \bar{\mcl{Z}}
\end{align}
where $F$ is the same as in \eqref{eq:J}. 
A globally optimal solution to this static optimization problem is denoted as $\bar z^\star$.
The set of optimal steady-state pairs is denoted by $\bar{\mcl{Z}}^\star$, i.e.,
\begin{equation}
\bar{\mcl{Z}}^\star =
 \left\{\bar z^\star \in \bar{\mcl{Z}}~|~\bar z^\star  \text{ is optimal in\,\eqref{eq:OPT_ss}}\right\}.
\end{equation}
Henceforth, we assume that $\bar{\mcl{Z}}^\star \not= \emptyset$.
The sets $\bar{\mcl{X}}$ and $\bar{\mcl{X}}^\star$, with $\bar{\mcl{X}}^\star \subseteq \bar{\mcl{X}} \subset \mcl{X}$, denote the projection of $\bar{\mcl{Z}} \subset \Rnx\times\Rnu$ onto the state space $\Rnx$ and the projection of $\bar{\mcl{Z}}^\star \subset \Rnx\times\Rnu$ onto $\Rnx$, respectively.

In the operation of dynamic processes, it is of major interest to know whether the best infinite-horizon performance can be achieved at the best steady state or via permanent transient operation. Optimal operation over an infinite horizon is defined similar to \cite{Gruene13a,Angeli12a} as follows.
\begin{defi}[Optimal operation at steady state] \label{def:S}
System~\eqref{eq:sys_def} is said to be optimally operated at steady state if there exists a $\bar z = (\bar x, \bar u)^T \in \bar{\mcl{Z}}$ such that, for any initial condition $x_0 \in \mcl{X}_0$ and any infinite-time admissible pair $z(\cdot, x_0)$,
we have
\begin{equation} \label{eq:optOptSS}
\underset{T\to\infty}{\liminf} \, J_T(x_0, u(\cdot)) \geq F(\bar z). 
~ \vspace{-4mm}
\end{equation}
  
\end{defi}
The following lemma follows trivially from the above. 
\begin{lemm}\label{lem:opt_ss_op}
If system~\eqref{eq:sys_def} is optimally operated at  $\bar z$, then $\bar z$ is an optimal solution to \eqref{eq:OPT_ss}.
 
\end{lemm}

\subsection{Turnpike Properties of OCPs}
Since there is no generally valid definition of turnpike properties of continuous-time OCPs, we propose a definition motivated by a turnpike result given in \cite{Carlson91}.  To this end, consider the placeholder variable 
$
\xi \in \left\{x, z\right\}, 
$
which, depending on the context, denotes the state or the input-state pair of \eqref{eq:sys_def}. Accordingly, $\xi^\star(\cdot, x_0)$ is either an optimal state trajectory $x^\star(\cdot)$ or an optimal pair $z^\star(\cdot)$. Likewise, 
$\bar \xi$ is either a steady state $\bar x$ or a steady-state pair $\bar z$. 
Using this placeholder variable, we define the set 
\begin{equation} \label{eq:Theta}
\Theta_{\xi,T}(\varepsilon) = \left\{t \in [0,T]: \left\| \xi^\star(t, x_0, u^\star(\cdot)) - \bar \xi\right\| > \varepsilon\right\}.
\end{equation}

\begin{figure}
\begin{center}
\includegraphics[width=0.4\textwidth]{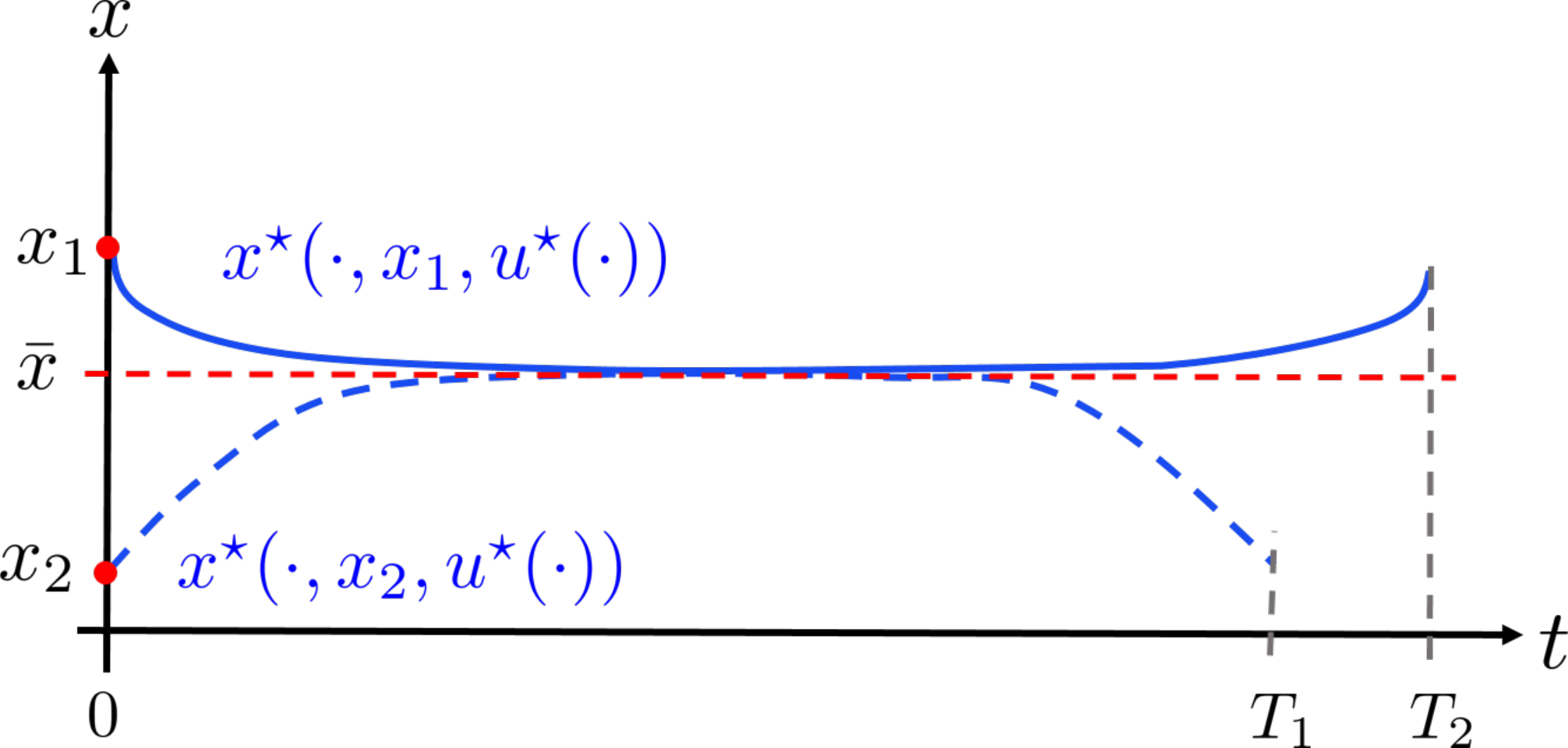}
\caption{Sketch of a state turnpike property. 
\label{fig:turnpike}}
\end{center}
\end{figure}

\begin{defi}[Turnpike property of $\OCP{x_0}$] \label{def:turnpike}
The solution pairs $z^\star(\cdot, x_0)$ of $\OCP{x_0}$ are said to have a \textnormal{$\{$state, input-state$\}$ turnpike} with respect to $\bar \xi  \in \left\{\bar x, \bar z\right\}$ 
if there exists a function $\nu_\xi:[0,\infty)\to [0,\infty]$ such that, for all $x_0 \in \mathcal{X}_0$ and all $T>0$, we have
\begin{equation}\label{eq:TP}
\mu\left[\Theta_{\xi,T}(\varepsilon)\right]  <\nu_\xi(\varepsilon)< \infty \quad \forall\: \varepsilon >0,
\end{equation}
where $\mu[\cdot]$ is the Lebesgue measure on the real line.\\
The pairs $z^\star(\cdot, x_0)$ of \eqref{eq:OCP} are said to have an \textnormal{exact $\{$state, input-state$\}$ turnpike} if Condition \eqref{eq:TP} also holds for $\varepsilon = 0$, i.e., 
\begin{equation} \label{eq:exactTP}
\mu\left[\Theta_{\xi,T}(0)\right]  <  \nu_\xi(0) < \infty. ~ 
\end{equation}
 \end{defi}
%
The turnpike property states that, for any initial condition $x_0$ and any horizon length $T>0$, the time that the optimal solutions of $\OCP{x_0}$ spend outside an $\varepsilon$-neighborhood of $\bar \xi$ is bounded by $\nu_\xi(\varepsilon)$, where $\nu_\xi(\varepsilon)$ is independent of the horizon length $T$. In other words, for sufficiently long horizons $T$, the optimal solutions have to enter any arbitrarily small $\varepsilon$-neighborhood of $\bar \xi$. In the case of a state turnpike, i.e. $\xi = x$, the solutions have to be close to the steady state $\bar x$. This situation is sketched in Fig. \ref{fig:turnpike}. In the case of an input-state turnpike, i.e. $\xi = z$, also the inputs have to approach any arbitrarily small $\varepsilon$-neighborhood of $\bar u$.
Note that, the steady state $\bar x$, respectively, the steady-state pair $\bar z$ approached by the optimal solutions is commonly referred to as \textit{the turnpike}.
According to Def. \ref{def:turnpike}, the turnpike has to be global in the sense that it is the same for all horizon lengths $T \geq 0$ and all $x_0 \in \mcl{X}_0$. Naturally, it also possible to have local turnpike properties. 
Here, for the sake of simplicity, we focus on global turnpike properties. 

If the stronger condition \eqref{eq:exactTP} holds, then, for sufficiently long horizons $T$, the optimal solutions have to enter the turnpike \textit{exactly} for some part of the horizon, i.e., the optimal solutions have to be at steady state for some part of the horizon. Hence, this case is denoted as exact turnpike, for an analysis  see \cite{epfl:faulwasser15a, kit:faulwasser17a}.


\vspace*{-2mm}
\begin{rema}[Extremal turnpikes]~\\
As shown in \cite{Trelat15a} for OCPs without input or path constraints, the adjoint variables also stay close to their optimal steady-state values whenever states and inputs stay close to the turnpike. It is straightforward to modify Def. \ref{def:turnpike} to include this third case. Without further elaboration, we suggest to denote this case as an \textit{extremal turnpike}.
\end{rema}
%
\begin{rema}[Turnpikes and reachability]\label{rem:turnpike_imp_reach}~\\
Def.~\ref{def:turnpike} implies that the turnpike steady state $\bar x$ is asymptotically reachable from all $x_0 \in \mathcal{X}_0$.  
\end{rema}

\subsection{Dissipativity}
Next, we briefly recall the definition of dissipativity with respect to a steady state \cite{Angeli12a}. We refer to \cite{Willems72a, Moylan14a} for further details on dissipativity. 
Let $w: \mcl{X}\times\mcl{U}\to \R$  be given by
\begin{equation} \label{eq:def_w}
w(x,u) := F(x,u) - F(\bar z),
\end{equation}
with $\bar z\in\bar{\mcl{Z}}$, and $F$ is the cost function in \eqref{eq:J} and  \eqref{eq:OPT_ss}. Let $\mcl{K}$ denote the set of functions of class $\mcl{K}$. 
\begin{defi}[Dissipativity wrt  a steady state]\label{def:diss}
System  \eqref{eq:sys_def} is said to be \textit{dissipative on $\mcl{Z}$ with respect to $\bar z  \in \bar{\mcl{Z}}$} if 
there exists a non-negative storage function\footnote{Note that the required properties of $S$ differ in different works: in \cite{epfl:faulwasser14e, Mueller14a} boundedness is assumed, while in \cite{Angeli12a} the storage $S$ can take real values instead of non-negative real values.} $S:\mcl{X} \to \R^+_0$ such that for all $x_0 \in \mcl{X}$, all $T\geq 0$ and all $u(\cdot) \in  \mcl{L}([0,T], \mcl{U})$ satisfying $x(t, x_0, u(\cdot)) \in \mcl{X}$ for all $t \in [0,T]$ we have
\begin{subequations} \label{eq:DI}
\begin{equation}\label{eq:diss}
S(x_T) - S(x_0) \leq \int_{0}^{T} w(x(t), u(t))\,\mathrm{d}t,
\end{equation}
where  $x_T = x(T, x_0, u(\cdot))$. \\
If, in addition, for some $\alpha \in \mcl{K}$ and $\xi \in \left\{x, z\right\}$,
\begin{multline} \label{eq:str_diss}
S(x_T) - S(x_0) \\ \leq \int_{0}^{T} -\alpha\left(\left\|\xi(t)-\bar \xi\right\|\right) + w(x(t), u(t))\,\mathrm{d}t
\end{multline}
\end{subequations}
then, 
\begin{enumerate}[(i)]
\item for $\xi = x$, system \eqref{eq:sys_def} is said to be \textnormal{strictly dissipative on $\mcl{Z}$ with respect to the steady state $\bar x$};  
\item for $\xi = z$, system \eqref{eq:sys_def} is said to be \textnormal{strictly dissipative on $\mcl{Z}$ with respect to the steady-state pair $\bar z$};
\end{enumerate}
\end{defi}
\begin{defi}[Dissipativity of $\OCP{x_0}$]~\\
If, for all  $x_0 \in \mcl{X}_0$, \eqref{eq:diss}---respectively \eqref{eq:str_diss}---is satisfied along any optimal pair of $\OCP{x_0}$,
 we say that  \emph{$\OCP{x_0}$ is dissipative---respectively strictly dissipative---with respect to $\bar \xi \in \{\bar x, \bar z\}$.} 
\end{defi}
 Note that in the non-strict case, one can show that dissipativity of system \eqref{eq:sys_def} and dissipativity of $\OCP{x_0}$ are equivalent. 
Furthermore, strict dissipativity of \eqref{eq:sys_def} implies strict dissipativity of $\OCP{x_0}$ but not vice versa.\footnote{Note that in \cite{Stieler14a} \emph{dissipativitiy of a system with respect to a steady state} is denoted as \emph{dissipativity of an OCP}.} Due to these dissipativity notions, $w$ in \eqref{eq:def_w} is called a \textit{supply rate}. 

In order to verify (strict) dissipativity, one has two possibilities: compute a storage function, or rely on converse dissipativity results. Next, we recall such a result.
To this end, consider the OCP 
\vspace*{-2mm}
\begin{subequations} \label{eq:OCP_diss}
\begin{align} 
\sup_{u(\cdot) \in  \mcl{L}([0,T], \mcl{U}),\,T }&  -\int_0^T  w(x(t), u(t))\,\mathrm{d}t \label{eq:OCP_diss_Sa}\\
&\hspace*{-1cm}\textrm{subject to }   \nonumber \\
\dot x &= f(x,u), \quad x(0) = x_0, \\
\forall t \in  [0,T]: \quad x(t) &\in \mcl{X}, \quad u(t) \in \mcl{U},\\
T&\geq 0, \label{eq:OCP_diss_T} \vspace*{-2mm}
\end{align}
\end{subequations}
which allows a free end time $T$ in \eqref{eq:OCP_diss_T}.
If $F(\bar z) = 0$ holds, then \eqref{eq:OCP_diss} is the free end-time version of $\OCP{x_0}$.
Let $S^a(x_0)$ denote the optimal value function of  \eqref{eq:OCP_diss}, which is also called the available storage. Observe that $S^a(x_0)\geq 0$, since $T = 0$ is  allowed. The following result states a necessary and sufficient condition for dissipativity. 
\vspace*{-2mm}
\begin{thm}[Willems 1972 \cite{Willems72a}]\label{thm:dissCond} ~\\
 System \eqref{eq:sys_def} is dissipative on $\mcl{Z}= \mcl{X}\times\mcl{U}$ if and only if $S^a(x_0)$ is finite for all $x_0\in\mcl{X}$.
  Moreover, if the system is dissipative with respect to the supply rate w, then $0 \le S^a(x_0)\le S(x_0)$ for \emph{any} storage function $S$ and the function $S^a$ itself is a storage function.         
\end{thm}
The proof can be obtained by straightforward modification of the classical proof from \cite{Willems72a}. The next corollary is directly implied by $0 \le S^a(x_0)\le S(x_0)$.
\vspace*{-2mm}
\begin{coro}
System \eqref{eq:sys_def} is dissipative with bounded non-negative storage function, if and only if, for all $x_0 \in \mcl{X}$, $S^a(x_0) \leq \hat S < \infty$.  
\end{coro}
\vspace*{-2mm}
\begin{rema}[Verifying strict dissipativity]\label{rema:str_diss}~\\
Note that in order to verifiy strict dissipativity on a compact set, one uses that Thm. \ref{thm:dissCond} applies to generic supply rates $w$, i.e. one swaps $w$ with $-\alpha + w$ in \eqref{eq:OCP_diss_Sa} and shows that, for at least one $\alpha\in\mcl{K}$, the available storage $S^a$ is finite.
\end{rema}
\vspace*{-2mm}
\begin{rema}[Strict dissipativity of $\OCP{x_0}$]\label{rema:dissOCP}~\\
To show strict dissipativity of $\OCP{x_0}$, one swaps $w$ with $-\alpha + w$ and "$u(\cdot) \in  \mcl{L}([0,T], \mcl{U})$" in \eqref{eq:OCP_diss_Sa} with "$u(\cdot)$ is optimal in~$\OCP{x_0}$". Again Thm. \ref{thm:dissCond} holds with straightforward modification to the proof in \cite{Willems72a}.
\end{rema}

\subsection{Problem Statement} \label{sec:results}
\begin{figure}[t]
\begin{center}
\includegraphics[width=0.49\textwidth]{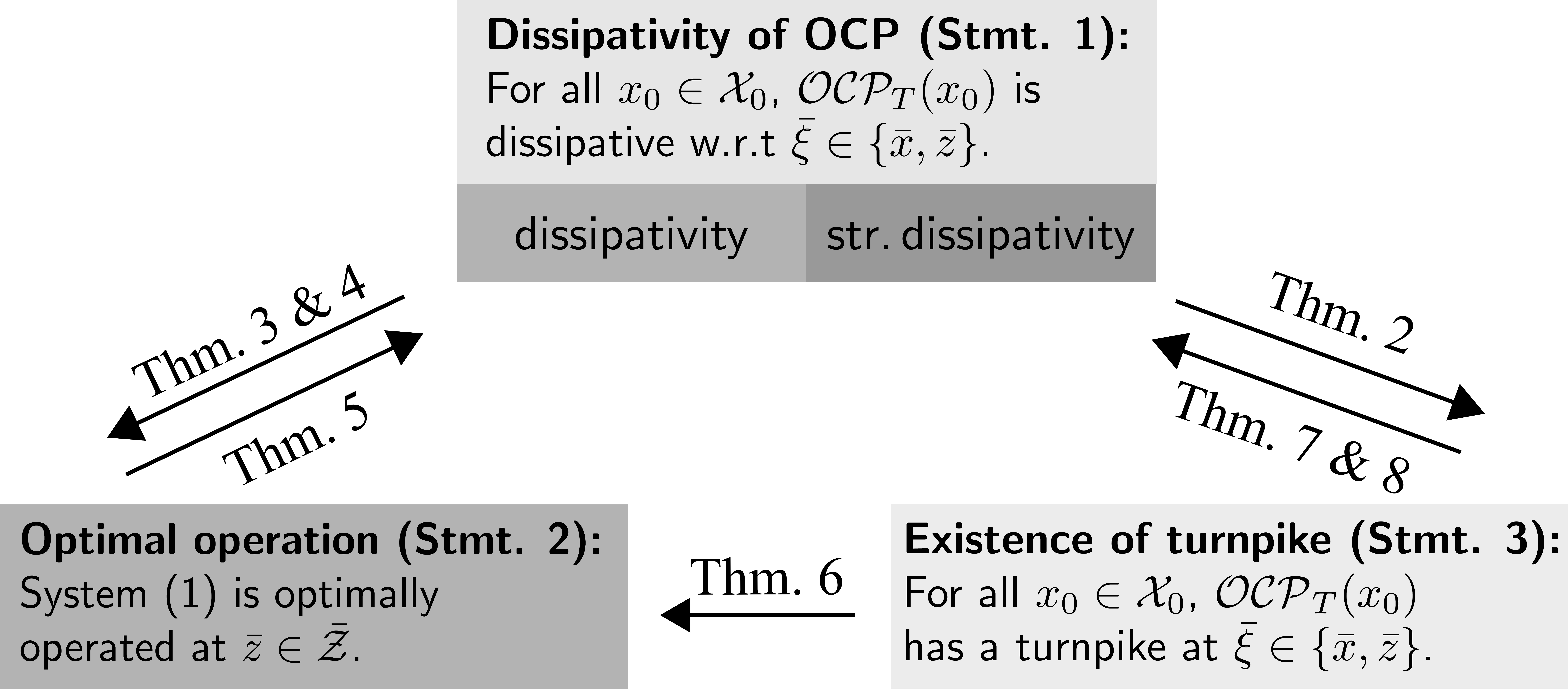}
\caption{Implications between dissipativity, optimal operation at steady state and turnpikes investigated in this paper. \label{fig:implications}}
\end{center}
\end{figure}

The main purpose of this paper is to establish links between the three following formal statements/assumptions:
\begin{stat} \label{stat:D}
For all $x_0 \in \mcl{X}_0$, $\OCP{x_0}$ is strictly dissipative with respect to $\bar \xi \in \{\bar x, \bar z\}$ and with bounded storage function.
\end{stat}
\vspace*{-3mm}
It is worth mentioning that if Stmt. \ref{stat:D} is true for $\bar \xi = \bar z$, then it also holds for $\bar \xi = \bar x$.  
\vspace*{-3mm}
\begin{stat} \label{stat:S}
System \eqref{eq:sys_def} is opt. operated at $\bar z^\star \in \bar{\mcl{Z}}^\star$.
 \end{stat}
\vspace*{-3mm}
\begin{stat} \label{stat:TP}
For all $x_0 \in \mcl{X}_0$, the optimal solutions of $\OCP{x_0}$ have a $\{$state, input-state$\}$ turnpike at $\bar \xi \in \{\bar x, \bar z\}$.
 \end{stat}
 \vspace*{-2mm}
Subsequently, we show that 
\begin{itemize}
\vspace*{-2mm}
\item Stmt. \ref{stat:D} $\Rightarrow$ Stmt. \ref{stat:TP} (Thm. \ref{thm:D_imp_TP});
\item Stmt. \ref{stat:D} $\Rightarrow$ Stmt. \ref{stat:S} (Thm. \ref{thm:D_imp_S});
\item Stmt. \ref{stat:TP} $\Rightarrow$ Stmt. \ref{stat:S} (Thm. \ref{thm:TP_imp_S});
\item Stmt. \ref{stat:S} $\Rightarrow$ Stmt. \ref{stat:D} (Thm. \ref{thm:S_imp_D});
\item Stmt. \ref{stat:TP} $\Rightarrow$ Stmt. \ref{stat:D} (Thm. \ref{thm:exTP_imp_D} and Thm. \ref{thm:TP_imp_D}).
\end{itemize}
 An overview of the results to be shown is sketched in Fig. \ref{fig:implications}.
 %
%

For some of these relations we will invoke a reachability assumption. 
\begin{ass}[Exponential reachability] \label{ass:reach}~\\
For all $x_0\in \mcl{X}_0$, there exists an optimal steady state pair $\bar z^\star\in \bar{\mcl{Z}}^\star$, an infinite-time admissible input $ u_\infty(\cdot, x_0) \in \mcl{L}([0,\infty), \mcl{U})$, and constants $c, \lambda \in \mathbb{R}^+$, independent of $x_0$, such that
\[
\left\| (x(t, x_0,  u_\infty(\cdot, x_0)), \,u_\infty(\cdot, x_0))^T - \bar z^\star\right\| \leq c e^{-\lambda t}. 
\vspace*{-4mm}
\]
 \end{ass}
This assumption is satisfied if one can steer the state to any small neighborhood of $ \bar x^\star$ in finite time, $ \bar x^\star$ corresponds to a   steady-state pair $( \bar x^\star,  \bar u^\star) \in \inte Z$, and \eqref{eq:sys_def} has a stabilizable Jacobian linearization at this point.


\section{Implications of Dissipativity} \label{sec:D}
\begin{thm}[Dissipativity $\Rightarrow$ turnpike]  \label{thm:D_imp_TP}~\\
Let Stmt.~\ref{stat:D}   hold for $\xi \in \{x, z\}$ and Ass.~\ref{ass:reach} holds for $\bar z^\star~\in~\bar{\mcl{Z}}$.
Then, Stmt. \ref{stat:TP} holds, i.e., $\OCP{x_0}$ has a turnpike at $\bar \xi^\star\in \{\bar x ^\star, \bar z^\star\}$.  
\end{thm}
\vspace*{-4mm}
\begin{pf}
We first consider $\xi =z$. Let $z^\star(\cdot,x_0)$ be an optimal solution to $\OCP{x_0}$. The integral dissipation inequality~(\ref{eq:str_diss}) gives
\begin{align*}
S(x^\star(T)) - S(x^\star(0))\le &  -\int_0^T \alpha(\| z^\star(t,x_0) - \bar z^\star \|)\,\mathrm{d}t \\ & + 
\int_0^T\hspace{-2mm} F(z^\star(t)) - F(\bar z^\star)\,\mathrm{d}t.
\end{align*}
By Ass.~\ref{ass:reach}, $\bar x^\star$ is exponentially reachable from every $x_0\in\mcl{X}_0$. Hence, the second integral is bounded from above by $K_F:=\frac{c}{\lambda}L_F < \infty$ independently of $T$, where $L_F$ is a Lipschitz constant of $F$ and $c,\lambda$ are from Ass.~\ref{ass:reach}. In addition, since $S$ is bounded, the lhs is bounded in absolute value, independently of $T$, by some $K_S < \infty$. Hence, we have
\begin{align*}
\int_0^T \alpha(\| z^\star(t,x_0) - \bar z^\star \|)\,\mathrm{d}t \le K_S +K_F < \infty.
\end{align*}
Using \eqref{eq:Theta} and that $\alpha \in \mcl{K}$, we obtain
\begin{equation*}
\int_0^T \alpha(\| z^\star(t,x_0) - \bar z^\star \|)\,\mathrm{d}t \ge 
 \alpha(\varepsilon)\mu[\Theta_{z,T}(\varepsilon)].
\end{equation*}
It follows that
$
\mu[\Theta_{z,T}(\varepsilon)] \le (K_S+K_F) / \alpha(\varepsilon) =: \nu_z(\varepsilon),
$
where $\nu_z(\varepsilon)$ does not depend on $T$.
The case $\xi =x$ can be proved easily along the same lines. \hfill~ \QED
\end{pf}
%
%
%
\begin{thm}[Dissipativity $\Rightarrow$ opt. operation at $\bar z^\star$] \label{thm:D_imp_S}
If Stmt.~\ref{stat:D} holds at $\bar z^\star$, then Stmt.~\ref{stat:S} holds, i.e., system \eqref{eq:sys_def} is optimally operated at the steady state $\bar z^\star$. 
\end{thm}
\begin{pf}
(By contradiction). Assume that there exists an infinite-time optimal pair $z^\star_\infty(\cdot, x_0)$ and a sequence $\{T_k\}_{k=1}^\infty$, with $T_{k+1}\ge T_k$ and $T_k\to\infty$, such that
\begin{equation}\label{eq:proofDimpSs^aux}
\lim_{k\to\infty} J_{T_k}(x_0, u^\star_\infty(\cdot)) \le  F(\bar z^\star) - \sigma
\end{equation}
for some $\sigma > 0$. Evaluating the dissipation inequality~\eqref{eq:diss} along $z^\star_\infty(\cdot, x_0)$ and dividing by $T_k$ gives
\[
\frac{1}{T_k}[S(x^\star_\infty(T_k)) - S(x^\star_\infty(0))] \le     J_{T_k}(x_0, u^\star_\infty(\cdot)) - F(\bar z^\star).
\]
Since the storage function $S$ is bounded, the lhs of the above inequality converges to zero for $T_k \to \infty$, 
whereas the rhs converges to $-\sigma< 0$, which is a contradiction. \QED
\end{pf}
\vspace*{-5mm}
In the proof of Thm.~\ref{thm:D_imp_S}, we never invoked the strict dissipativity term $\alpha(\|\xi - \bar \xi^\star\|)$. Hence, the following stronger statement holds.
\begin{thm}\label{thm:D_imp_SII}~\\
For all $x_ 0 \in \mcl{X}_0$, let  $\OCP{x_0}$ be dissipative (not necessarily strictly) with bounded storage and with respect to the steady-state pair~$\bar z^\star$. Then, Statement~\ref{stat:S} holds, i.e., the system \eqref{eq:sys_def} is optimally operated at the steady state~$\bar z^\star$.  
\end{thm}

\section{Implications of Optimal Operation \newline
at Steady State} \label{sec:SS}
In order to discuss the implications of optimal operation at steady state, we will extend a discrete-time result given in \cite{Mueller14a} to the continuous-time setting.
To this end,
consider the set
\begin{subequations} \label{eq:sets}
\begin{multline} \label{eq:steerable_set}
\mcl{C}(\bar x^\star, T) := \left\{x_0 \in  \mcl{X} \,|\,\exists u(\cdot) \in \mcl{L}([0,T], \mcl{U}), \right.\\ 
\hspace*{.9cm}\, \forall t\in[0,T]:  x(t, x_0, u(\cdot)) \in \mcl{X}, \\\left.
\, x(T, x_0, u(\cdot)) = \bar x^\star\right\},
\end{multline}
which is the set of initial conditions $x_0 \in \mcl{X}$ that can be steered, in some finite time $T \in (0,\infty)$, by means of an admissible input, to $\bar x^\star \in \mcl{X}$.
Likewise, we define the set
\begin{multline} \label{eq:reachable_set}
\mcl{R}(\bar x^\star, T):= \left\{x_T\in  \mcl{X} \,|\,\exists u(\cdot) \in \mcl{L}([0,T], \mcl{U}), \right.\\ 
\hspace*{.9cm}\, \forall t\in[0,T]:  x(t, \bar x^\star, u(\cdot)) \in \mcl{X}, \\
\left.\, x(T, \bar x^\star, u(\cdot)) = x_T\right\}, 
\end{multline}
\end{subequations}
which containts all states that can be reached, in some finite time $T \in (0,\infty)$, by means of an admissible input, starting from $\bar x^\star$. 
Since, by construction, $\bar x^\star$ is contained in both sets, we have $\mcl{C}(\bar x^\star, T) \cap \mcl{R}(\bar x^\star, T) \neq \emptyset$.
Now, let 
\begin{multline} \label{eq:X_T}
\mcl{X}_T :=  \left\{x_0 \in  \mcl{X} \,|\,\forall t\in[0,T]: \,\exists u(\cdot) \in \mcl{L}([0,T], \mcl{U})  \right.\\ \left.
\hspace*{-.6cm}\, x(t, x_0, u(\cdot)) \in \mcl{C}(\bar x^\star, T) \cap \mcl{R}(\bar x^\star, T) \right\} 
\end{multline}
be the set of initial conditions $x_0$, for which (i) there exists a corresponding admissible pair $z(\cdot, x_0, u(\cdot))$, and (ii) the inclusion $x(t, x_0, u(\cdot))\in \mcl{C}(\bar x^\star, T) \cap \mcl{R}(\bar x^\star, T)$ holds for all $t\in[0,T]$. Subsequently, we use the shorthand notation $\mcl{Z}_T :=\mcl{X}_T\times\mcl{U} $.

\begin{thm}[Opt. operation at $\bar z^\star \Rightarrow$ dissipativity] \label{thm:S_imp_D}
If Stmt. \ref{stat:S} holds, then, for any $T \geq 0$, Stmt. \ref{stat:D} holds with $\mcl{Z}$ replaced by  $\mcl{Z}_T $, i.e.,  system \eqref{eq:sys_def} is dissipative on $\mcl{Z}_T $.
\end{thm}
 The proof follows along the same lines as the discrete-time version \cite[Thm. 4]{Mueller14a} and is thus omitted. 
\begin{coro} 
If Stmt. \ref{stat:S} holds, then, for any $T \geq 0$, Stmt. \ref{stat:D} holds with $\mcl{X}_0 = \mcl{X}_T $, i.e.,  $\OCP{x_0}$ is dissipative.
\end{coro}

\section{Converse Turnpike Results} \label{sec:T}
After discussing the implications of dissipativity and optimal operation at steady state, we now turn to converse turnpike results.

\begin{thm}[Turnpike $\Rightarrow$  opt. operation at $\bar z$] \label{thm:TP_imp_S}~\\
Let Stmt. \ref{stat:TP} hold for $\bar \xi \in \{\bar x, \bar z\}$. If either $\bar \xi = \bar z$ or  $\bar\xi = \bar x$ and
\begin{equation} \label{eq:cond_ThmConvTP}
\frac{\partial F}{\partial u} = 0, \quad \forall  (x,u) \in \mcl{Z},	
\end{equation}
 then Stmt. \ref{stat:S} holds, i.e., \eqref{eq:sys_def} is optimally operated at ~$\bar{z}$. 
 \end{thm}
\vspace*{-5mm}
\begin{pf}
We first consider the case  $\bar \xi = \bar z$. 
Fix $x_0\in \mathcal{X}_0$ and, for contradiction, assume that there exists an infinite-time admissible pair $z_\infty(\cdot, x_0)$ and a sequence $\{T_k\}_{k=1}^\infty$ with $T_{k+1}\ge T_k$ and $T_k\to\infty$ such that
\begin{equation}\label{eq:proof_aux}
\lim_{k\to\infty} J_{T_k}(x_0,u_\infty(\cdot)) \le  F(\bar z) - \sigma
\end{equation}
for some $\sigma > 0$. 
Next, observe that the turnpike property and Lipschitz continuity of $F(x,u)$ imply that there exists a function $\gamma_z$, independent of $T$, such that
\[ 
\mu[\Omega_{z,T}(\varepsilon)] < \gamma_z(\varepsilon),
\] 
with
$
\Omega_{z,T}(\varepsilon) := \{t\in [0,T] \mid F(z_T^\star(t, x_0)) - F(\bar z)| > \varepsilon \}$.
Set $m :=\min_{z\in\mathcal{Z}}F(z)$ and $\bar \Omega := [0,T_k]\setminus \Omega_{z,T_k}(\varepsilon) $. Then, for arbitrary $\varepsilon > 0$, we have
\begin{align*}
\int_0^{T_k}F(z_{T_k}^\star(t))\,\mathrm{d}t &= \int_{\Omega_{z,T}(\varepsilon)} \hspace{-0.5cm}  F(z_{T_k}^\star(t))\,\mathrm{d}t + \int_{\bar \Omega}F(z_{T_k}^\star(t))\,\mathrm{d}t  \\
 \hspace{-2.4cm} \ge m\mu[\Omega_{z,T}(\varepsilon)] &+  \int_{\bar \Omega} \hspace{-0.2cm}F(\bar{z})-\varepsilon\,dt \\
= m\mu[\Omega_{z,T}(\varepsilon)] &+  \int_0^{T_k} \hspace{-0.2cm}F(\bar{z})-\varepsilon\, dt - \int_{\Omega_{z,T}(\varepsilon)} \hspace{-0.4cm} F(\bar{z})-\varepsilon\,\mathrm{d}t \\
= m\mu[\Omega_{z,T}(\varepsilon)] &+ T_k(F(\bar{z})-\varepsilon) - \mu[\Omega_{z,T}(\varepsilon)](F(\bar{z})-\varepsilon).
\end{align*}
Since $\mu[\Omega_{z,T}(\varepsilon] < \gamma_z(\varepsilon)$, independently of $T_k$, dividing by $T_k$ and letting $k\to\infty$ gives
\[
\lim_{k\to\infty} \frac{1}{T_k}\int_0^{T_k}F(z_{T_k}^\star(t))\,\mathrm{d}t  \ge F(\bar{z}) - \varepsilon .
\]
Selecting $\varepsilon < \sigma$ leads to a contradiction to~(\ref{eq:proof_aux}), since the pair $z_\infty(\cdot, x_0)$ truncated 
to  $[0,T_k]$ is admissible for $\OCP{x_0}$ for any $T_k \ge 0$.

Recall that, for $\bar \xi = \bar x$,  \eqref{eq:cond_ThmConvTP} implies that $F$ does not depend on $u$. 
Hence, swapping $z$ with $x$ in the derivations above, proves the assertion for  $\bar \xi = \bar x$. \QED
\end{pf}
\vspace*{-5mm}
Thm.~\ref{thm:TP_imp_S} and Lem.~\ref{lem:opt_ss_op} lead to the following corollary. 
\begin{coro}[Turnpikes are opt. steady states] \label{coro2} 
Let Stmt. \ref{stat:TP} hold for $\bar \xi \in \{\bar x, \bar z\}$ and assume that  the conditions of Theorem~\ref{thm:TP_imp_S} hold. Then, for $\bar \xi = \bar z$,
the turnpike $\bar z$ is optimal in~(\ref{eq:OPT_ss}), i.e., $\bar z = \bar z^\star$; and, for $\bar \xi = \bar x$, there exists a $\bar u\in\mcl{U}$ such that $(\bar{x},\bar u)$  is optimal in~(\ref{eq:OPT_ss}).  
\end{coro}
\vspace*{-2mm} Corollary \ref{coro2} is a consequence of the turnpike definition used in this paper, which implies asymptotic reachability of $\bar{x}$ from all $x_0\in \mcl{X}_0$ (see Remark~\ref{rem:turnpike_imp_reach}). If, instead, a local definition of turnpike is used, analogous local results can be established.  
%

At this point, one may wonder whether or not Stmts. \ref{stat:D}--\ref{stat:TP} are equivalent. 
In the view of Thm. \ref{thm:D_imp_S} \& \ref{thm:S_imp_D} this boils down to the question of
how large the gap between strict and non-strict dissipativity, i.e., between \eqref{eq:diss} and \eqref{eq:str_diss}, is.
One intuitive way to close this gap is assuming that $F$ has a unique minimizer $\bar{z}^\star$ over $\mathcal{Z}$, and this unique minimizer is a steady state, i.e., $\bar{z}^\star \in \bar{\mathcal{Z}}$. This, however, would imply to restrict the class of cost functions significantly. 
Another approach, pursued in \cite{Gruene16a} in a discrete-time setting, requires local controllability close to $\bar z$ and turnpike-like behavior of nearly optimal solutions. 
Next, we follow a different route and present two distinct set of conditions guaranteeing that the existence of a turnpike implies dissipativity. 
\begin{thm}[Exact turnpike $\Rightarrow$ dissipativity]\label{thm:exTP_imp_D}~\\
Let Stmt. \ref{stat:TP}  hold with $\mcl{X}_0 = \mcl{X}$, and let the turnpike be exact and of the input-state type ($\xi = z$). 
Then,
\vspace*{-3mm}
\begin{enumerate}[(i)]
\item  system \eqref{eq:sys_def} is dissipative with respect to $\bar z^\star$ on $\mcl{Z}$ and with bounded storage function, and
\item 
$\OCP{x_0}$ is strictly dissipative with respect to $\bar z^\star$.  
\end{enumerate}
\end{thm}
\vspace*{-5mm}
\begin{pf}
\underline{Part (i):} Since, for all $T \geq 0$, the optimal pairs $z^\star(\cdot, x_0)$ show turnpike behavior, we have 
\begin{multline*}
|s^a(x_0)|  := \left|\int_0^T w(z^\star(t, x_0))\,\mathrm{d}t\right| \\
\leq \int_{\Theta_{z, T}(\varepsilon)}\hspace*{-5mm} |w(z^\star(t, x_0)) |\,\mathrm{d}t  
+ \int_{[0,T]\setminus \Theta_{z, T}(\varepsilon) } \hspace*{-8mm} |w(z^\star(t, x_0)) |\,\mathrm{d}t. 
\end{multline*}
Due to Lipschitz continuity of $F$, for all $T \geq 0$, all $x_0 \in \mcl{X}$, and all $\varepsilon >0$, the second integral on the rhs is bounded from above by $T L_F\varepsilon$, where $L_F$ is a Lipschitz constant of $F$.
Hence, the last inequality leads to
\begin{equation*}
|s^a(x_0)|
\leq \mu[\Theta_{z, T}(\varepsilon)] \hat w + T L_F\varepsilon, 
\end{equation*}
where $\hat w := \sup_{z \in\mcl{Z}} |w(z)| < \infty$.
The turnpike is exact. We may set $\varepsilon  = 0$ and obtain
$
|s^a(x_0)|
\leq \nu_z(0) \hat w < \infty$.
Hence, for all $T\geq 0$ and all $x_0 \in \mcl{X}_0$, the supremum in \eqref{eq:OCP_diss} is finite.  
Using Thm. \ref{thm:dissCond}, we conclude that \eqref{eq:sys_def} is dissipative  with bounded storage function
on $\mcl{Z}$. 

\vspace*{-2mm}
\underline{Part (ii):} Choose any $\alpha \in \mcl{K}$, and split the time horizon into $\Theta_{z, T}(\varepsilon)$ and $[0,T]\setminus\Theta_{z, T}(\varepsilon)$. Similar to part (i), this leads to
\begin{align*}
|s^a(x_0)| &:= \left|\int_0^T \hspace*{-2mm} -\alpha(\|z^\star(t, x_0) - \bar z^\star\|) + w(z^\star(t, x_0))\,\mathrm{d}t \right|  \\
&\leq \mu[\Theta_{z, T}(\varepsilon)] \left(\hat\alpha+ \hat w\right) + T (\alpha(\varepsilon)+  L_F\varepsilon), 
\end{align*}
where $\hat\alpha := \sup_{z \in \mcl{Z}}\, \alpha(\|z - \bar z^\star\|) < \infty$.
Setting $\varepsilon  = 0$ allows concluding that, for all $T \geq 0$ and all $x_0 \in \mcl{X}_0$,
%
$| s^a(x_0)| < \infty$.
Note that Thm. \ref{thm:dissCond} also holds if one restricts the class of considered input signals in \eqref{eq:OCP_diss} (Rem. \ref{rema:dissOCP}).  
Hence, we conclude that along optimal pairs of $\OCP{x_0}$ the strict inequality \eqref{eq:str_diss} holds for the chosen $\alpha \in \mathcal{K}$. 
\QED
\end{pf}
\vspace*{-5mm}
Let $\mcl{B}_\rho(\bar z)$ denote an open ball of radius $\rho$ centered at $\bar z$.
\vspace*{-2mm}
\begin{ass}[$\bar z^\star$  is local minimizer of $F$ on $\mcl{Z}$] \label{ass:loc_bnd}
 There exists a constant $\rho >0$ and $\alpha_\rho \in \mcl{K}$ such that
 \begin{equation} \label{eq:loc_bnd}
 \alpha_\rho(\|z - \bar{z} \|) \leq F(z) - F(\bar z), \quad \forall z \in \mcl{B}_\rho(\bar z) \cap \mcl{Z}. 
 \end{equation} 
\end{ass}
%
%
\vspace*{-4mm}
\begin{thm}[Turnpike $\Rightarrow$ strict dissipativity] \label{thm:TP_imp_D}
Suppose that 
 Ass. \ref{ass:loc_bnd} and Stmt. \ref{stat:TP} hold,  and let the turnpike be of the input-state type ($\xi = z$). \\
Then,  there exists a bounded storage function $S$ depending on $\alpha_\rho$, s.t.  $\OCP{x_0}$ is strictly dissipative with respect to $\bar z^\star$. 
\vspace*{-4mm} 
\end{thm}
\vspace*{-2mm}
\begin{pf}
Consider $\alpha_\rho \in \mcl{K}$ from Ass.  \ref{ass:loc_bnd}
and the integral
\begin{equation} \label{eq:strDI_opt}
s^a(x_0)= \int_0^T  \alpha_\rho(\|z^\star(t, x_0) - \bar z^\star\|) - w(z^\star(t, x_0))\,\mathrm{d}t
\end{equation}
evaluated along optimal pairs $z^\star(\cdot, x_0)$ of $\OCP{x_0}$. 
We split the time horizon into $\Theta_{z, T}(\varepsilon)$ and $[0,T]\setminus\Theta_{z, T}(\varepsilon)$. 
Evaluating the rhs of \eqref{eq:strDI_opt} on $\Theta_{z, T}(\varepsilon)$ yields
\begin{subequations} \label{eq:bnd1}
\begin{multline}
\int_{\Theta_{z, T}(\varepsilon)} \hspace{-2mm} \alpha_\rho(\|z^\star(t, x_0) - \bar z^\star\|) - w(z^\star(t, x_0))\,\mathrm{d}t \\
 \leq \mu[\Theta_{z, T}(\varepsilon)](\hat\alpha+\hat w) 
\end{multline}
where $\hat\alpha_\rho:= \sup_{z \in \mcl{Z}}\, \alpha_ \rho(\|z - \bar z^\star\|) < \infty$, and $\hat w$ is as defined in proof of Thm. \ref{thm:exTP_imp_D}. The turnpike property of $\OCP{x_0}$ implies, for all $T>0$ and all $x_0 \in \mcl{X}_0$, that
\begin{equation}
\mu[\Theta_{z, T}(\varepsilon)](\hat\alpha+\hat w) \leq \nu_z(\varepsilon)(\hat\alpha+\hat w)  < \infty.
\end{equation}
\end{subequations}
To evaluate the rhs of \eqref{eq:strDI_opt} on  $[0,T]\setminus\Theta_{z, T}(\varepsilon)$, recall that the turnpike of $\OCP{x_0}$
implies that for $T$ sufficiently large, we have
$
\|z^\star(t, x_0)-\bar z\|\leq \rho,\quad \forall t \in [0,T]\setminus\Theta_{z, T}(\varepsilon),
$
where $\rho$ is from Ass. \ref{ass:loc_bnd}. 
Using \eqref{eq:loc_bnd} we obtain
\begin{equation} \label{eq:bnd2}
\int_{[0,T]\setminus\Theta_{z, T}(\varepsilon)} \hspace{-3mm}  \alpha_\rho(\|z^\star(t, x_0) - \bar z^\star\|) - w(z^\star(t, x_0))\,\mathrm{d}t \leq 0.
\end{equation}
Combining \eqref{eq:bnd1} and \eqref{eq:bnd2} yields
$
 s^a(x_0) <  \nu_z(\varepsilon)(\hat\alpha+\hat w) < \infty.
$
Recalling Rem. \ref{rema:dissOCP}, we conclude that along optimal pairs  \eqref{eq:str_diss} holds for $\alpha_\rho$. \QED
\end{pf}

\section{Example: Chemical Reactor} \label{sec:example}

We consider the example of a continuously stirred chemical  reactor.
A model of the reactor, including the concentration of species $A$ and $B$, $c_A, c_B$ in mol/l and the reactor temperature $ \vartheta$ in $^\circ$C as state variables, reads
\begin{subequations} \label{eq:sys_vanVuss}
\begin{align}
\dot c_A & = -r_A(c_A, \vartheta) + (c_{in}-c_A)u_1 \\
\dot c_B & =\phantom{-} r_B(c_A, c_B,  \vartheta)  -c_Bu_1 \\
\dot  \vartheta  & = \phantom{-}h(c_A, c_B,  \vartheta) + \alpha(u_2- \vartheta) + ( \vartheta_{in}- \vartheta)u_1, 
\end{align}
\end{subequations}
where $r_A 	   =k_1c_A +k_3c_A^2, \quad r_B  = k_1c_A-k_2c_B$,\quad $h = -\delta( k_1 c_A \Delta H_{AB} + k_2c_B \Delta H_{BC}+k_3c_A^2 \Delta H_{AD})$
 and $k_i = \phantom{-}k_{i0}e^{\frac{-E_i}{ \vartheta+ \vartheta_0}}, \quad i = 1,2,3$.
 The system parameters can be found in \cite{Rothfuss96b}. 
The states and inputs are subject to the constraints $
c_A \in [0, 6] \frac{mol}{l}, c_B \in [0, 4]\frac{mol}{l},  \vartheta \in [70, 150]^\circ C$ and $u_1 \in [3, 35]\frac{1}{h}, u_2 \in [0, 200]^\circ C$.
We consider the problem of maximizing the production rate of $c_B$; thus $F$ in \eqref{eq:J} and \eqref{eq:OPT_ss} is
$F(c_B, u_1) = -\beta c_Bu_1, \beta >0.$
%
%
%
%
To numerically verify dissipativity, we approximate the exponential term $k_i(\theta)$ by its fourth-order Taylor expansion at $\vartheta = 110^\circ$C.
This way the system dynamics become polynomial and a polynomial storage function can be sought using sum-of-squares (SOS) programming. 
Indeed, by choosing a quadratic function $\alpha := \bar{\alpha} \|x - \bar{x}^\star \|^2$, where $\bar{\alpha} > 0 $ is an optimization variable, all data in \eqref{eq:str_diss} become polynomial. We solve the following problem
\begin{subequations} \label{eq:OPT_sos}
\begin{align}
&\underset{\bar{\alpha} \in [0,1], S(\cdot)\in \mathbb{R}_d[x]}{\max} \, \bar{\alpha}   \\
\textrm{subject to } &\forall (x,u)\in \mcl{X}\times \mcl{U}\nonumber \\ 
 w(x,u) &- \bar{\alpha}\|x - \bar x^\star \|^2 -\frac{\partial S}{\partial x}f(x,u) \geq 0, \label{eq:sos}
\end{align}
\end{subequations}
where $\mathbb{R}_d[x]$ denotes the vector space of polynomials of fixed degree $d$. The non-negativity constraint~(\ref{eq:sos}) is replaced by a sufficient SOS constraint; here we use the standard Putinar condition~\cite{Putinar93} imposed for each of the four vertices of $\mcl{U}$ separately (this is possible since $u$ enters the dynamics affinely and $\mcl{U}$ is convex). Instead of the integral inequality~\eqref{eq:str_diss} we consider its differential counter part in \eqref{eq:sos}.
This leads to a semidefinite programming problem, which is solved using SeDuMi~\cite{sedumi}. Details of sum-of-squares programming are omitted for brevity; see \cite{lasserreBook,Ebenbauer09a}. The strict dissipativity constant $\bar{\alpha}$ is constrained to $[0,1]$ for numerical reasons.
The optimal steady state $\bar x^\star = [2.1756, 1.1049, 128.53]^T$, $\bar u^\star = [ 35, 142.76]^T$ is computed using an SQP-method; its global optimality is verified using Gloptipoly~\cite{gloptipoly3}.
Seeking a polynomial storage function of degree five via \eqref{eq:OPT_sos} is feasible with $\bar{\alpha} = 1$, thus verifying strict dissipativity with respect to $\bar x^\star$ on $\mcl{X}\times\mcl{U}$.\\
\begin{figure}[t]
\begin{center}
\includegraphics[width=0.495\textwidth]{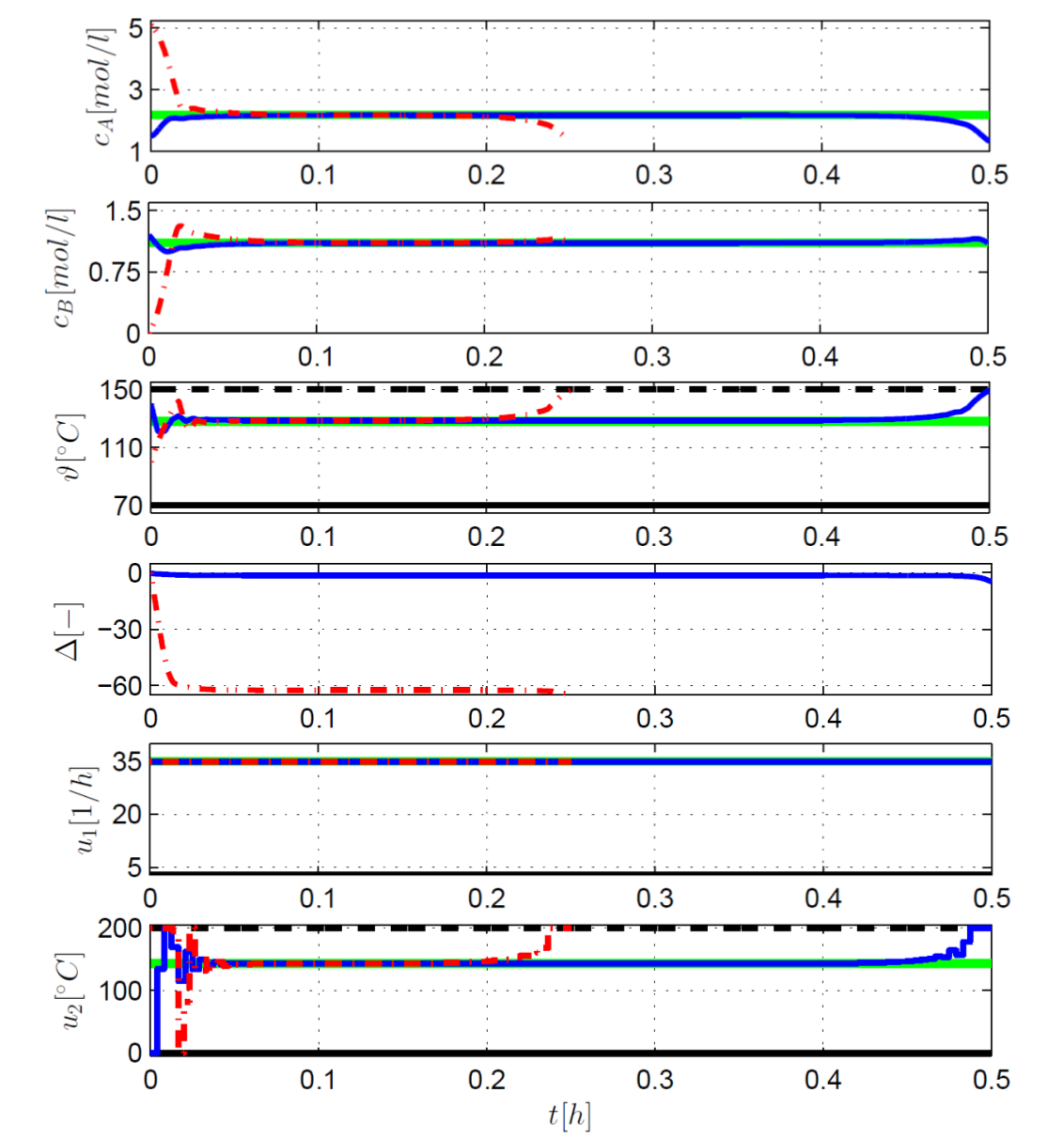}  \\
\caption{Simulation results for  \eqref{eq:sys_vanVuss}: 
blue -- $T = 0.5$; dash-dot red -- $ T = 0.25$; green -- optimal steady state; black and dash-dot black -- constraints. \label{fig:example}}
\end{center}
\end{figure}
To illustrate turnpike behavior, we solve \eqref{eq:OCP} as given above
 for two initial conditions with a piecewise-constant input parametrization with \cite{Houska11a}. 
For the initial condition $(c_A(0), c_B(0), \vartheta(0))  = (1.5, 1.2, 140)$, we consider an optimization horizon of $T = 0.5$; for $(c_A(0), c_B(0), \vartheta(0))  = (5.1, 0, 100)$, we use $T =0.25$. 
The plots in the upper part of Fig.~\ref{fig:example} show the state trajectories $c_A, c_B$ and $\vartheta$ and their optimal steady-state values.  Clearly, the optimal solutions exhibit the turnpike property. 
The lower part of Fig. \ref{fig:example} depicts the inputs $u_1, u_2$ (last two plots). Note that the input $u_1$ is always at its upper limit. The third plot from the bottom illustrates the strict dissipation inequality~\eqref{eq:str_diss}. To this end, we define  the residual
\[
\Delta(t) = S(x(\tau))\Big|_{0}^{t} 
-\int_0^t  \hspace*{-2mm} w(x(\tau),u(\tau)) -\bar\alpha\|x(\tau)-\bar x^\star \|\,\mathrm{d}\tau 
\]
where $\bar\alpha\|x-\bar x^\star \|$, $w(x,u)$ and the storage $S$ are the ones computed in \eqref{eq:OPT_sos}.


\section{Summary and Conclusions}
Before concluding this paper, it is worth summing up the main points and highlighting subtle differences between the presented results. 
Thms. \ref{thm:D_imp_TP}--\ref{thm:D_imp_SII}  rely on strict dissipativity of $\OCP{x_0}$, i.e. they require (strict) dissipativity to hold along all optimal solutions. Taking this into account, it is evident that equivalence of strict dissipativity of $\OCP{x_0}$ (Stmt. \ref{stat:D}) and a turnpike property of $\OCP{x_0}$ (Stmt. \ref{stat:TP}) is guaranteed under mild conditions, cf. 
 Ass. \ref{ass:reach} \& \ref{ass:loc_bnd} in Thm. \ref{thm:TP_imp_D}.
 It seems to be the generality of optimal operation at steady state (Stmt. \ref{stat:S}) which in search for equivalence conditions requires stronger assumptions such as local controllability of the turnpike steady state and turnpike-like behavior of near optimal solutions, cf.  \cite{Gruene16a}. 
 

For continuous-time optimal control problems, this paper has investigated the relationship between dissipativity properties, optimal operation at steady state, and turnpike properties. We extended discrete-time results to show that
dissipativity of $\OCP{x_0}$ implies (a) optimal operation at steady state and (b) a turnpike property of optimal solutions. 
On top, we derived novel converse turnpike results showing that under mild assumptions a tunrpike in $\OCP{x_0}$ implies dissipativity. Finally, we commented on sufficient conditions guaranteeing the equivalence of the properties.


\bibliographystyle{plain}        

\end{document}